\documentclass[12pt]{iopart}

\usepackage{color}   
\usepackage{graphicx}
\usepackage{epsfig,bm}

\begin{document}

\title{Exploring Early Parton Momentum Distribution with the Ridge
from the Near-Side Jet}

\author{Cheuk-Yin Wong}

\address{Physics Division, Oak Ridge National Laboratory,
Oak Ridge, TN 37830, U.S.A.}
\ead{wongc@ornl.gov}
\begin{abstract}
In a central nucleus-nucleus collision at high-energies, medium
partons kicked by a near-side jet acquire a momentum along the jet
direction and subsequently materialize as the observed ridge
particles.  They carry direct information on the early parton momentum
distribution which can be extracted by using the ridge data for
central AuAu collisions at $\sqrt{s_{NN}}=200$ GeV.  The extracted
parton momentum distribution has a thermal-like transverse momentum
distribution but a non-Gaussian, relatively flat rapidity distribution
at mid-rapidity with sharp kinematic boundaries at large rapidities
that depend on the transverse momentum.
\end{abstract}

\nopagebreak


In central high-energy heavy-ion collisions, the state of the parton
medium during the early stage of a nucleus-nucleus collision is an
important physical quantity.  It furnishes information for the
investigation of the mechanism of parton production in the early
stages of the collision of two heavy nuclei.  It also provides the
initial data for the evolution of the system toward the state of
quark-gluon plasma.  Not much is know about the early state of the
partons from direct experimental measurements.

Recently, the STAR Collaboration observed a $\Delta \phi$-$\Delta
\eta$ correlation of particles associated with a near-side jet in
central AuAu collisions at $\sqrt{s_{NN}}=200$ GeV at RHIC, where
$\Delta \phi$ and $\Delta \eta$ are the azimuthal angle and
pseudorapidity differences relative to a high-$p_t$ trigger particle
\cite{Ada05,Put07,Wan07,Bie07}. The near-side correlations can
be decomposed into a ``jet'' component with fragmentation and radiation
products at $(\Delta \phi, \Delta \eta)$$
\sim$(0,0), and a ``ridge'' component at $\Delta\phi$$ \sim$0 with a
ridge structure in $\Delta \eta$.

While many theoretical models have been proposed
\cite{Hwa07,Won07,Won07a}, a momentum kick model was presented to
describe the ridge phenomenon \cite{Won07,Won07a}. The model assumes
that a near-side jet occurs near the surface and it kicks medium
partons, loses energy along its way, and fragments into the trigger
particle and other fragmentation products in the ``jet'' component.
The kicked medium partons, each of which acquires a momentum kick
${\bf q}$ from the near-side jet, materialize by parton-hadron duality
as ridge particles.  The ridge particle momentum distribution is
related to the initial momentum distribution by a momentum shift.  We
can therefore extract the initial parton momentum distribution from
the ridge data for central AuAu collisions at $\sqrt{s_{NN}}=200$ GeV
obtained by the STAR Collaboration \cite{Ada05,Put07,Wan07}.

We parametrize the normalized initial parton momentum distribution as
\begin{eqnarray}
\label{dis2}
~~~~~~~~
\frac{dF}{ p_{t}dp_{t}dy d\phi}&=&
A_{\rm ridge} (1-x)^a 
\frac{ e^ { -\sqrt{m^2+p_{t}^2}/T }} {\sqrt{m_d^2+p_{t}^2}},
\end{eqnarray}
where $x$=${\sqrt{m^2+p_{t}^2}} e^{|y|-y_b} /{m_b} \le 1$, $A_{\rm
ridge}$ is a normalization constant defined by $\int d{\bf p} dF/d{\bf
p} =1$, $a$ is the rapidity fall-off parameter, $y_b$ is beam parton
rapidity, $m_b$ is the beam parton mass, $m$ = $m_\pi$, and $m_d$ is a
mass parameter introduced to give a better description of low-$p_t$
ridge data.  For lack of a definitive determination, we set $y_b$
equal to the nucleon rapidity $y_N$ and $m_b$ equal to $m$, pending
future definitive measurements of the ridge boundary.  The observed
ridge distribution is the final parton momentum distribution after
jet-parton collisions, scaled by the average number of charged kicked
partons per trigger, $\langle N \rangle$, that may be modified
by a ridge attenuation factor $f_R$.

\begin{figure} [h]
\hspace{0.0in}
\includegraphics[angle=0,scale=0.40]{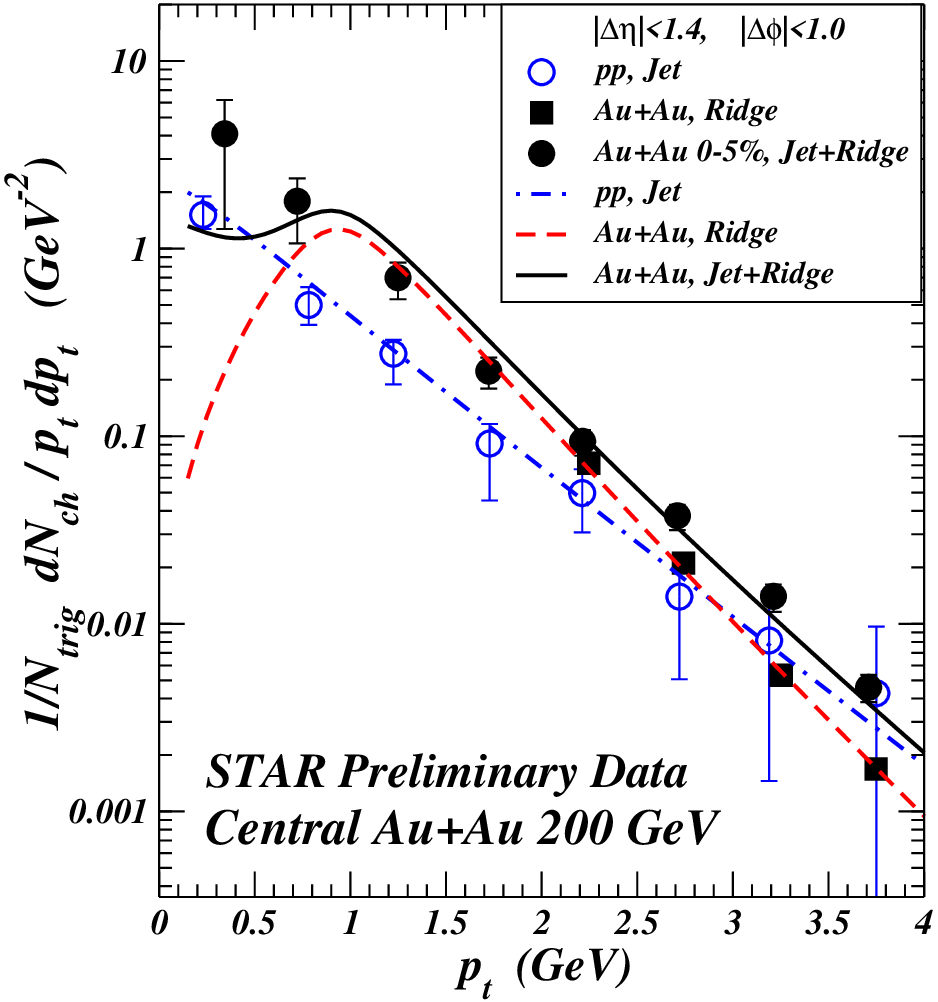}
\vspace*{0.0cm}
\end{figure}

\vspace{-8.0cm}
\begin{figure} [h]
\hspace{2.9in}
\includegraphics[angle=0,scale=0.40]{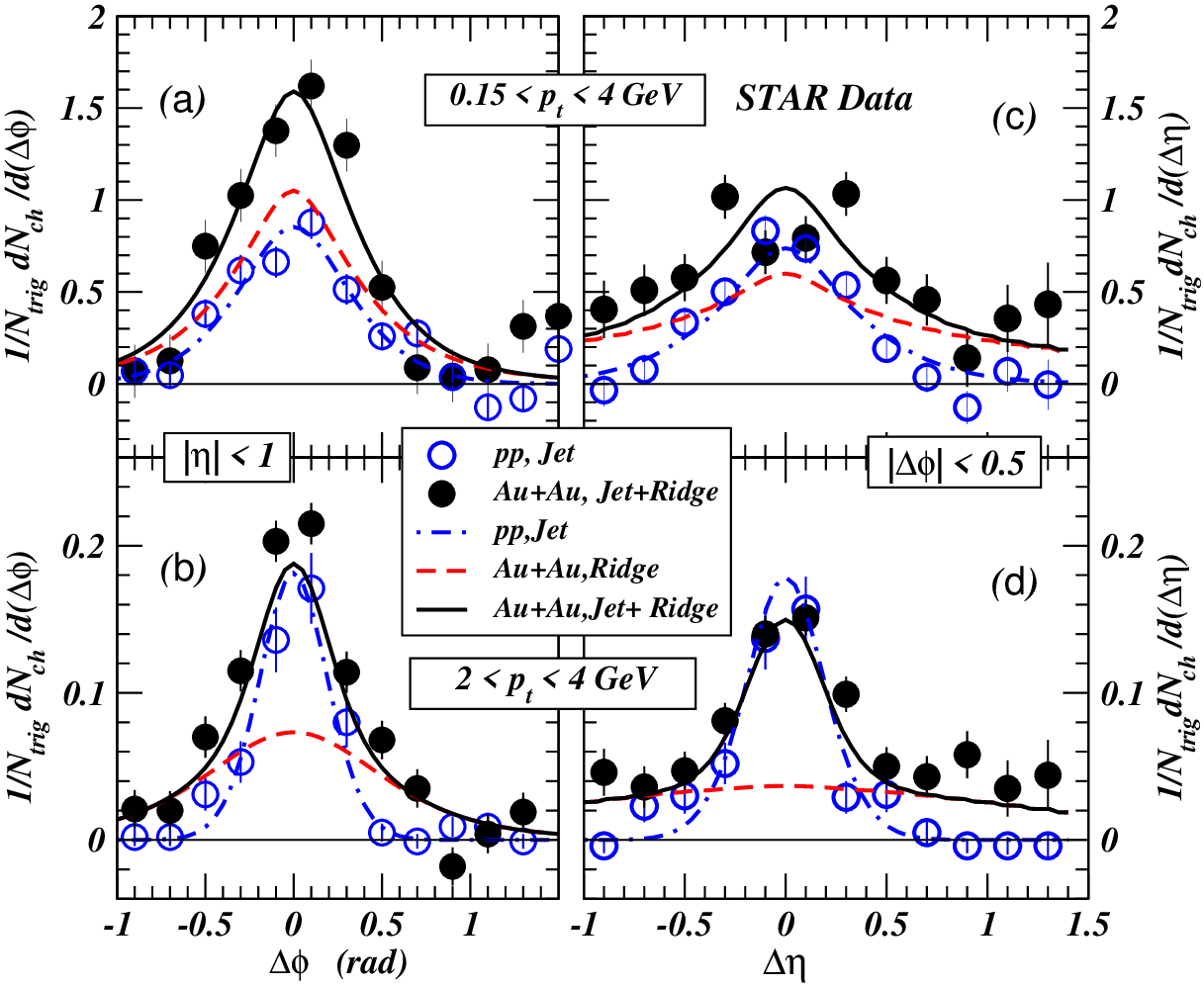}
\end{figure}


\vspace*{-0.4cm} \hangafter=-3 \hangindent=-9.0cm {\noindent{\bf
Fig.~1~} Transverse momentum distribution of associated particles in
$pp$ and central AuAu collisions.}

\vspace*{-1.6cm} \hangafter=-3 \hangindent=7.6cm {\noindent {\bf
Fig.~2~} Azimuthal angular and pseudorapidity distributions in $pp$ and
central AuAu collisions.}

\vspace*{0.2cm} To infer the associated ridge particle yield from
experimental data, we need to know the jet component in central AuAu
collisions.  The scaling relation between the fragments in the jet
component and the trigger particle \cite{Put07} allows us to treat the
AuAu jet component per trigger as a $pp$ near-side jet distribution,
attenuated by a semi-empirical attenuation factor $f_J=0.63$
\cite{Won07a}. The experimental $pp$ near-side jet data, shown as open
circles in Figs.\ 1 and 2, can be described by the dash-dot curves in
these figures obtained from
\begin{eqnarray}
\label{jetfun}
\fl
~~~~~
\frac { dN_{\rm jet}^{pp}} {p_t dp_t\, d\Delta \eta\, d\Delta \phi}
\!\!&=& N_{\rm jet}
\frac{\exp\{(m-\sqrt{m^2+p_t^2})/T_{\rm jet}\}} {T_{\rm jet}(m+T_{\rm jet})}
\frac{\exp\{{- {[(\Delta \phi)^2+(\Delta \eta)^2]}/{2\sigma_\phi^2} }\}}
      {2\pi\sigma_\phi^2},
\end{eqnarray}
where 
$\sigma_\phi$=$\sigma_{\phi 0}\,{m_a}/{\sqrt{m_a^2+p_t^2}}$,
$\sigma_{\phi 0}$=0.5, $m_a$=1.1 GeV,
$N_{\rm jet}$=0.75, and $T_{\rm jet}$=0.55 GeV.

Theoretical evaluation of both the jet component and the ridge
component for central AuAu collisions leads to the total yield of
associated particles.  A self-consistent comparison of the momentum
kick model results with experimental data in Figs.\ 1, 2, and 3 then
allows us to search for the initial parton momentum distribution.  We
find that the totality of the STAR associated particle data
\cite{Ada05,Put07, Wan07} from $p_t=0.15$ GeV to 4 GeV and $|\eta|$
from 0 up to 3.9 in central AuAu collisions at $\sqrt{s_{NN}}=200$
GeV, can be described by the average momentum kick per jet-parton
collision $|{\bf q}|=1$ GeV, the average effective numbers of kicked
partons per jet $f_R \langle N \rangle = 4$, and the
normalized initial parton momentum distribution Eq.\ (\ref{dis2}) with
\begin{eqnarray}
\label{bestset}
~~~~~~~~
a=0.5, T=0.50 {\rm ~GeV,~and~}
m_d=1 {\rm ~GeV}.  
\end{eqnarray}
Fig.\ 1 shows good agreement between theoretical $dN_{\rm ch}/N_{\rm
trig}p_t dp_t$ results with experimental charged hadron yields.  Note
that the theoretical ridge $dN_{\rm ch}/N_{\rm trig}p_t dp_t$ (the
dashed curve) has a peak at $p_t\sim |{\bf q}| \sim 1$ GeV, as a
result of the momentum kick.  In Fig.\ 2, comparison of theoretical
and experimental associated particle data indicates general agreement
over azimuthal angles [Figs. 2(a) and 2(b)] and over pseudorapidities
[Figs. 2(c) and 2(d)], for both $0.15 < p_t < 4 $ GeV [Figs. 2(a) and
2(c)] and $2 < p_t < 4 $ GeV [Figs. 2(b) and 2(d)].  The forward
rapidity data in Fig. 3 have large uncertainties and the value of
$a=0.5$ gives reasonable agreement with experiment.

\begin{figure} [h]
\hspace{0.3in}
\includegraphics[angle=0,scale=0.30]{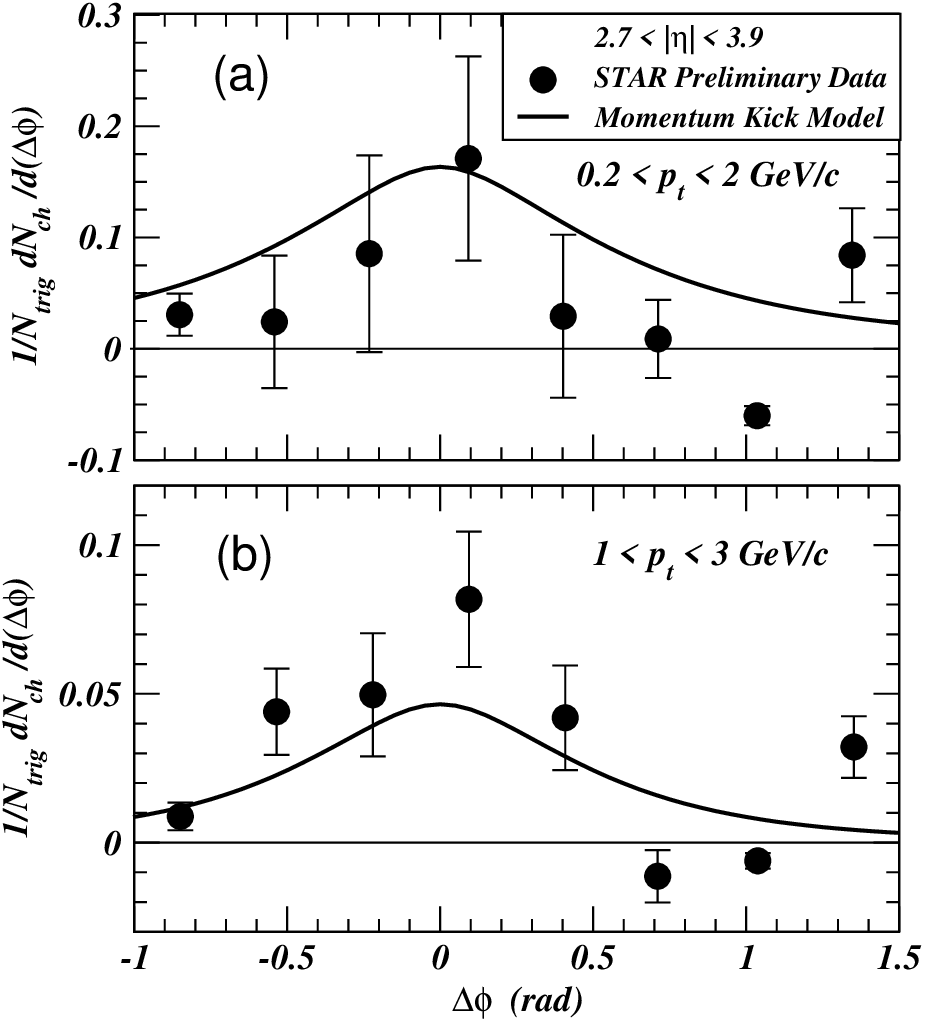}
\end{figure}

\vspace{-6.3cm}
\begin{figure} [h]
\hspace{2.9in}
\includegraphics[angle=0,scale=0.43]{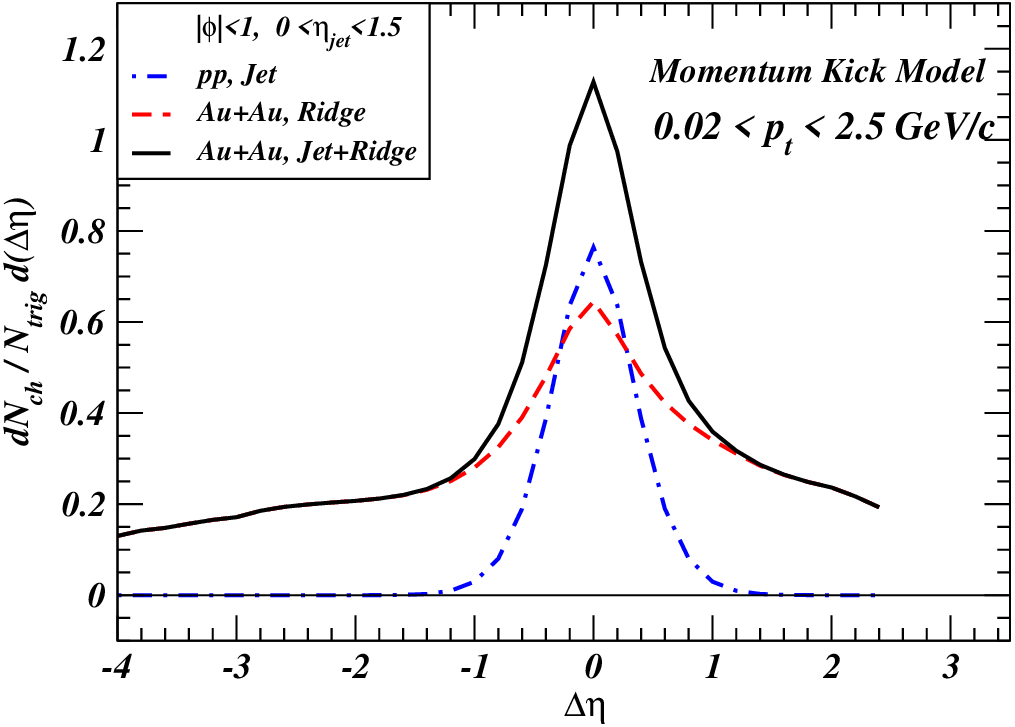}
\end{figure}

\vspace*{-0.5cm}
\hangafter=-3 \hangindent=-9.0cm {\noindent{\bf {Fig.~3~}} Azimuthal
distributions of associated particles in forward rapidities for
central AuAu collisions.}

\vspace*{-1.6cm} \hangafter=-3 \hangindent=7.4cm {\noindent {\bf
Fig.~4~} Theoretical predictions of associated particle pseudorapidity
distribution within the acceptance region of the PHOBOS detector.}

\vspace*{0.4cm} Using the parameters we have extracted from the STAR
ridge data for central AuAu collisions at $\sqrt{s}=200$ GeV, we can
predict the pseudorapidity distribution of associated particles for
the PHOBOS experimental acceptance defined by $\Delta \phi \le$1,
0$<\eta_{\rm trig}<$1.5, 0.02$<p_t<$2.5 GeV.  The total associated
particle yield is shown as the solid curve in Fig. 4.  The $pp$ jet
yield and the associated ridge particle yield are shown as the
dash-dot and the dashed curves respectively.  The result has been
corrected for $\Delta \eta$ acceptance.  The present prediction up to
large $|\Delta \eta|$ was found to agree well with experimental
measurements obtained by the PHOBOS Collaboration \cite{Wen08}.

\hangafter=9 \hangindent=3.2in The distribution (\ref{dis2}) with
parameters of Eq.\ (\ref{bestset}) gives the normalized initial parton
momentum distribution at the moment of jet-parton collision.  We show
this distribution in Figs.\ 5.  It cannot be separated as the product
of two independent distributions of $p_t$ and $y$.  In Fig. 5(a), the
momentum distribution for $y=0$ and high $p_t$ has a slope parameter
$T$ that is intermediate between that of the jet and the inclusive
particles.  This indicates that partons at the moment of jet-parton
collision are at an intermediate stage of dynamical equilibration.
The distribution as a function of $p_t$ does not change much as $y$
increases from 0 to 2.  For $y=3$, the maximum value of $p_t$ is 1.54
GeV and the distribution changes significantly as the kinematic
boundary is approached.  For $y=4$, the boundary of $p_t$ is located
at 0.55 GeV.

\vspace*{-0.6cm}
\begin{figure} [h]
\includegraphics[angle=0,scale=0.40]{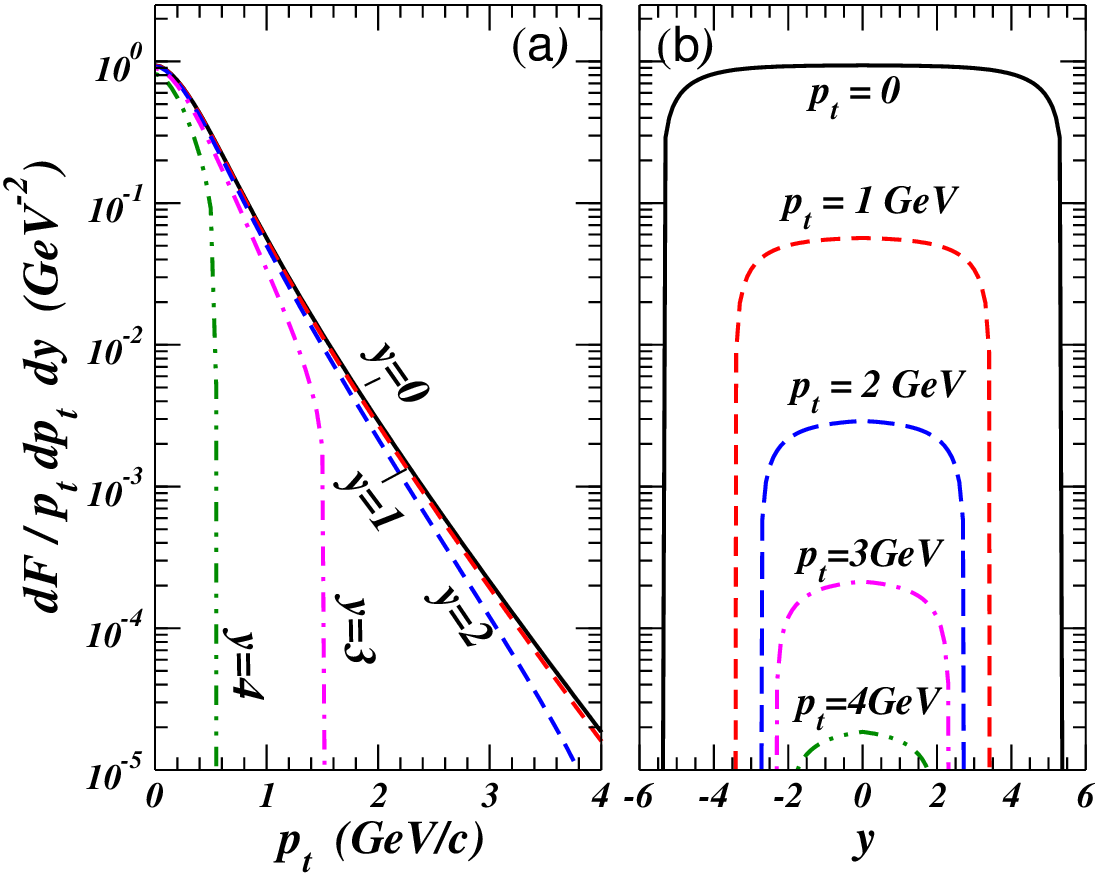}
\end{figure}

\vspace*{-0.4cm}
\hangafter=-3
\hangindent=-8.cm
\noindent{\bf Fig.~5~}
Initial parton momentum distribution
at the moment of jet-parton collisions.

\vspace*{-6.9cm} \hangafter=-11 \hangindent=3.2in In Fig.\ 5(b), the
momentum distribution as a function of $y$ for a fixed $p_t$ is
essentially flat at mid-rapidity and it extends to a maximum value of
$|y|_{\rm max}$ that depends on $p_t$.  The distribution decreases
rapidly as it approaches the kinematic limit and covers a smaller
allowed region of $y$ as $p_t$ increases.  The flat rapidity
distribution at mid-rapidity for a fixed $p_t$ gives rise to the ridge
structure that is observed experimentally.

\vspace*{-0.0cm} \hangafter=-2 \hangindent=3.2in In conclusion, the
application of the momentum kick model for ridge particles associated
with a near-side jet in central AuAu collisions at $\sqrt{s_{NN}}=200$
GeV allows us to extract unique and valuable information on the medium
parton momentum distribution at the moment of jet-parton collision.
In the process, we also find the average momentum kick per jet-parton
collision $|{\bf q}|$, and the average effective numbers of kicked
medium partons per near-side jet $f_R\langle N \rangle$, in
these collisions.  They provide important empirical data for future
investigations on the dynamics of parton production, parton evolution,
and jet momentum loss.

\end{document}